\documentclass[11pt,twoside]{article}
\usepackage{asp2010}

\resetcounters

\begin{document}

\title{Are magnetic OB stars more prone to mixing? Still an unsettled issue}
\author{Thierry Morel\affil{Institut d'Astrophysique et de G\'eophysique, Universit\'e de Li\`ege, Belgium}}

\begin{abstract}
We review our knowledge of the mixing properties of magnetic OB stars and discuss whether the observational data presently available support, as predicted by some theoretical models, the idea that magnetic phenomena favour the transport of the chemical elements. A (likely statistical) relationship between enhanced mixing and the existence of a field has been emerging over the last few years. As discussed in this contribution, however, a clear answer to this question is presently hampered by the lack of large and well-defined samples of magnetic and non-magnetic stars.
\end{abstract}

\section{Introduction}
The difficulties facing evolutionary models incorporating rotational mixing to reproduce the CNO abundance properties of some massive stars \citep[e.g.,][]{brott} suggest that some physical mechanisms that could significantly affect the amount of mixing experienced by single OB stars are still not (properly) accounted for in models. In this context, it is important to determine the surface chemical composition of well-established magnetic stars in the solar neighbourhood and to examine whether it differs in any way from that of stars without evidence for a field. Since the last review specifically dedicated to this topic \citep{morel11}, there have been a number of advances both on the theoretical and on the observational fronts. Contrary to one might have hoped, however, the present status of the field is arguably more confusing and clear conclusions about the impact that magnetic fields may have on mixing in O- and early B-type stars cannot be drawn yet. This unsatisfactory situation is in large part due to the difficulties in defining clear-cut samples of magnetic and non-magnetic stars and the hotly debated controversy surrounding the field detection in several objects. This pressing issue, and other recent developments in the field, are discussed in the following.

\section{Are magnetic OB stars more mixed?}
\subsection{New clues from theoretical models}
From the theoretical side, \cite{meynet} recently presented evolutionary models taking magnetic braking through a rotation-aligned, dipolar field into account. In the limit of strong differential rotation, these calculations show that it can significantly spin down stars with strong fields and lead to the appearance of slowly-rotating, yet nitrogen-rich main-sequence stars, as observed in different environments \citep[e.g.,][]{hunter, morel08}. However, the predictions are very sensitive to the poorly-known redistribution of the angular momentum in the stellar interior. When solid-body rotation is enforced, the decrease of the rotational velocity is very rapid. In addition, shear instabilities are suppressed and the efficiency of the transport of the chemical elements is low. As a result, very small nitrogen enrichments at the surface can be expected during the main-sequence phase. A determination of the internal rotation profile of some magnetic, pulsating stars (e.g., $\beta$ Cep) would be extremely valuable \citep[see, e.g.,][]{aerts03}, but this information is unfortunately not currently available.  

\subsection{New clues from abundance studies}
The CNO abundance properties of several magnetic stars have been investigated in a number of studies \citep[e.g.,][]{gies, kilian, morel08}, but a homogeneous determination for a large sample of B stars that include several candidate or well-established magnetic ones has recently been carried out by \citet{przybilla11}. This work confirms that several magnetic dwarfs, such as $\zeta$ Cas, $\beta$ Cep or $\tau$ \nolinebreak Sco, do present a large nitrogen overabundance at their surface (up to a factor 4). As discussed by \citet{morel08}, the low rotation rate of these stars is firmly established, either from the occurrence of phase-locked spectral variations that can be attributed to rotational modulation or from theoretical modelling in the pulsating variables.

In an effort to increase the number of magnetic stars with a detailed chemical composition, we have recently conducted a non-LTE abundance study of two stars detected as part of the MiMeS survey \citep{wade_alt}, namely HD 66665 and $\sigma$ Lup. Both stars have a relatively strong field, which is believed to magnetically confine the stellar wind \citep{petit, henrichs}. Assuming the radius ($R$ = 4.8$\pm$0.5 R$_{\odot}$) and very well constrained rotational period (3.02 d) proposed by \citet{henrichs}, one obtains that $\sigma$ Lup has an equatorial velocity in the range 70--90 km s$^{-1}$. HD 66665 is also an intrinsically slow rotator based on its narrow-lined nature ($v\sin i$ $<$ 10 km s$^{-1}$) and inclination with respect to the line of sight, $i$, which is greater than $\sim$15$^\circ$ \citep{petit}. In short, $\log g$ is determined from fitting the collisionally-broadened wings of the Balmer lines, $T_{\rm eff}$ from ionisation balance of various species (primarily Ne and Si) and the microturbulence, $\xi$, from requiring the abundances yielded by the O \,{\sc ii} features to be independent of their strength. The abundances are computed using Kurucz atmospheric models, an updated version of the non-LTE line-formation codes DETAIL/SURFACE \citep{giddings, butler} and classical curve-of-growth techniques. We made use of a high-resolution FIES spectrum acquired in the framework of the `fast-track service programme' of the NOT and an archival FEROS spectrum for HD 66665 and $\sigma$ Lup, respectively. Preliminary results are presented in Table 1. Although there is no indication for a contamination of the surface layers by core-processed material in two other OB stars detected by MiMeS (NGC 2244 \#201 and HD 57682; \citealt{morel11}), the high nitrogen-over-carbon abundance ratio ([N/C]) observed in HD 66665 and $\sigma$ Lup is instead strongly indicative of deep mixing.\footnote{The [N/O] ratio is, however, solar within the errors in the case of HD 66665. Although the origin of this inconsistency is currently unclear and needs to be investigated, the [N/C] ratio can be regarded as a more robust proxy of a nitrogen excess and is considered here as the primary mixing indicator.} This illustrates the diversity of the mixing properties of these four MiMeS stars, despite having relatively similar properties in terms of mass, evolutionary status, rotational velocity and field strength. The LTE abundance study of $\sigma$ Lup carried out by Henrichs et al. \citetext{in preparation} also supports a strong nitrogen overabundance. Sinusoidal modulations of the equivalent widths of several spectral lines are also reported and ascribed to the rotational modulation of abundance spots at the surface. This raises the interesting issue of the incidence of microscopic diffusion processes in early B-type stars in the presence of a strong field. Although we only have a snapshot spectrum at our disposal and are hence unable to investigate this aspect further, we note that it is very unlikely to substantially modify our conclusions regarding the N-rich status of this star: one of the three N \,{\sc ii} lines investigated shows small variations with a semi amplitude of 3--4\%, while none of the C \,{\sc ii} and O \,{\sc ii} lines under consideration exhibit significant changes.

\begin{table}
\caption{Preliminary atmospheric parameters and CNO elemental abundances of HD 66665 and $\sigma$ Lup (on a scale in which $\log \epsilon$[H]=12). The results of other studies in the literature (\citealt{petit}; Henrichs et al., in preparation) and those obtained for $\tau$ Sco using exactly the same tools and techniques as in the present work are shown for comparison \citep{hubrig08}. The number of lines used is given in brackets. A blank indicates that no value was determined. The solar [N/C] and [N/O] ratios are --0.60$\pm$0.08 and --0.86$\pm$0.08 dex, respectively \citep{asplund}.}
\label{table_abundances}
\vspace*{-0.5cm}
\begin{center}
\hspace*{-0.7cm}
{\scriptsize
\begin{tabular}{l|cc|cc|c} \hline
                        & \multicolumn{2}{c|}{HD 66665 (B0.5 V)}          & \multicolumn{2}{c|}{$\sigma$ Lup (B1.5 V)}          & $\tau$ Sco (B0.2 V)\\
                        & This study          & \citet{petit}    & This study           & Henrichs et al. \citetext{in prep.}    & Hubrig et al. (2008)\\\hline
$T_{\rm eff}$ (K)       & 27000$\pm$1000      & 28500$\pm$1000   & 25000$\pm$1000       & 23000$^a$           & 31500$\pm$1000\\
$\log g$ (cgs)          & 3.75$\pm$0.15       & 3.9$\pm$0.1      & 4.20$\pm$0.15        & 4.0$^a$             & 4.05$\pm$0.15\\
$\xi$ (km s$^{-1}$)     & 3$^{+2}_{-3}$       &                  & 3$\pm$3              &                     & 2$\pm$2\\ 
$v\sin i$ (km s$^{-1}$) & $<$10               & $<$10            & 68$\pm$3             & 68$\pm$6            & 8$\pm$2\\
$\log \epsilon$(C)      & 8.03$\pm$0.13 (7)   &                  & 8.18$\pm$0.06 (3)    &  7.90               & 8.19$\pm$0.14 (15)\\
$\log \epsilon$(N)      & 7.88$\pm$0.15 (32)  &                  & 8.30$\pm$0.17 (21)   &  8.30               & 8.15$\pm$0.20 (35)\\
$\log \epsilon$(O)      & 8.63$\pm$0.13 (33)  &                  & 8.54$\pm$0.26 (19)   &  8.75               & 8.62$\pm$0.20 (42)\\
${\rm [N/C]}$           & --0.15$\pm$0.17     &                  &  +0.12$\pm$0.19      & +0.40               & --0.04$\pm$0.25\\
${\rm [N/O]}$           & --0.75$\pm$0.16     &                  & --0.23$\pm$0.19      & --0.45              & --0.47$\pm$0.29\\\hline
\end{tabular}
}
\end{center}
\scriptsize{$^a$: Taken from the non-LTE study of \citet{levenhagen}.}
\end{table}

\subsection{Concerns about the reliability of the spectropolarimetric measurements}
Several stars with a tentative field detection or for which more observations were warranted have been re-observed over the past two years using different instrumentation and reduction/analysis techniques. \citet{hubrig11a, hubrig11b} obtained numerous field measurements for a sample of B stars suspected to various degrees to host a field using FORS at the VLT. They provided convincing evidence in the {\it CoRoT} target V1449 \nolinebreak Aql and in four other B stars for a sinusoidal modulation of the longitudinal field measurements with periods of a few days and full amplitudes amounting to $\sim$200 G (and even up to $\sim$1 kG for V1449 Aql). These data are consistent with a simple dipole field tilted with respect to the rotational axis. Interestingly, the rotation period obtained for V1449 \nolinebreak Aql under these assumptions (13.9 d) is consistent with the value independently obtained by \citet{aerts11} from a seismic analysis. However, a null detection was surprisingly reported by Shultz et al. \citetext{these proceedings} using ESPaDOnS at CFHT for three of the stars discussed by \citet{hubrig11b}. Even more troublesome is the fact that the individual FORS measurements can differ by up to several hundred Gauss with respect to those quoted by \citet{hubrig11b} when reduced using different procedures. This controversy clearly needs to be solved before firmly establishing the magnetic status of these stars. Until then, it may be fair to assume that $\alpha$ Pyx, 15 CMa and 33 Eri are candidate magnetic stars only. We will further assume in the following that stars investigated and detected by only one of the two groups are truly magnetic, which is another limitation to keep in mind.

\section{Discussion}
\begin{figure}[h!]
\includegraphics[width=11.8cm]{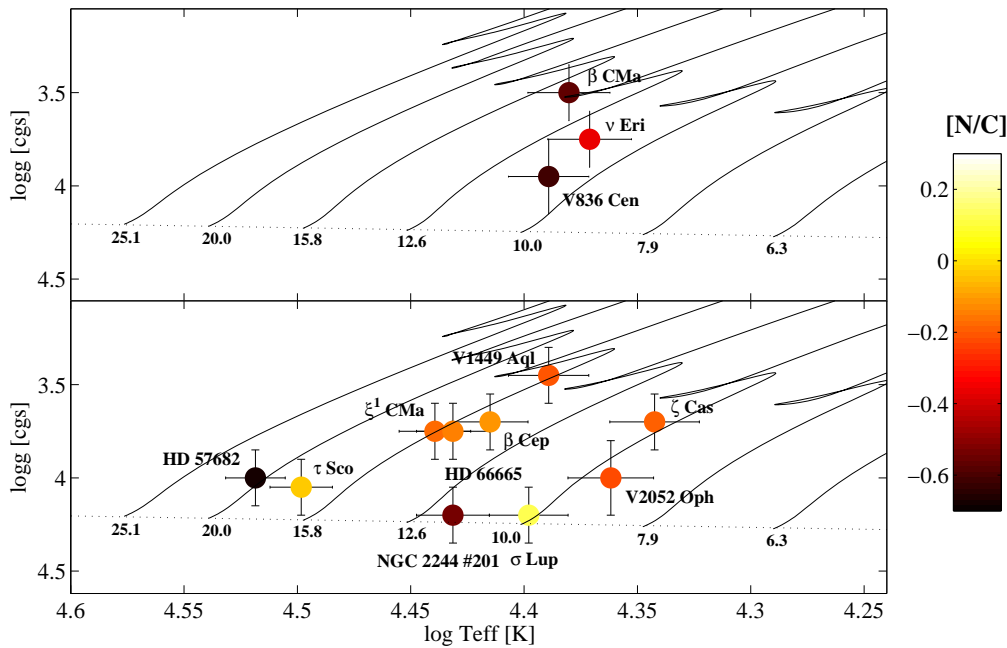}
\caption{Position in the $\log T_{\rm eff}$-$\log g$ plane of the OB stars without ({\it top panel}) and with ({\it bottom panel}) a magnetic field detection. The logarithmic N to C abundance ratio, [N/C], is colour coded (abundance data from \citealt{briquet_alt, morel08, morel11} and this study). Evolutionary tracks from \citet{claret} for masses ranging from 6.3 to 25.1 M$_{\odot}$ are overplotted.\label{fig_teff_logg}}
\end{figure}

Figure \ref{fig_teff_logg} shows the positions of the late O and early B stars with a field (un)detection that can be regarded as secure to a high level of confidence. Fields have also been detected in a number of more massive O stars (especially in those belonging to the Of?p class), but they generally lack a quantitative determination of the CNO abundances. The tendency for the magnetic objects to be more nitrogen rich and hence to display stronger evidence for deep mixing has previously been discussed by \citet{morel11}. This still holds true despite having considered two new objects and, more importantly, re-assessed the magnetic status of some of these stars. However, this conclusion is now based on a quite limited number of objects and, as a result, lies on much weaker grounds. On the other hand, it can be noticed that magnetic stars sharing similar properties can have drastically different surface abundance properties (e.g., HD 57682 and  $\tau$ \nolinebreak Sco). This suggests that other parameters might play a role in the amount of mixing experienced. New spectropolarimetric observations and, above all, reaching a reasonable level of agreement between the measurements of the different groups is badly needed before a coherent picture can hopefully be drawn. 

What can instead be regarded as undisputable is the existence of a population of slowly-rotating, unevolved B stars with evidence for CNO-cycled material at their surface. The prototype of this class of objects or, at least, the object with the best-known properties is $\tau$ Sco (keeping in mind that its field topology might turn out to be unusually complex). It is close to the ZAMS, is a truly slow rotator (equatorial rotational velocity as low as 6 km s$^{-1}$; \citealt{donati}) and yet displays the clear signature of deep mixing: [N/C]=--0.17$\pm$0.18 \citep{kilian}, --0.14$\pm$0.18 \citep{przybilla08} and --0.04$\pm$0.21 dex \citep{hubrig08}. Constructing models able to simultaneously reproduce these a priori conflicting properties constitutes a challenge, but promises to provide important clues concerning some physical mechanisms acting in massive stars. 

\acknowledgements The author acknowledges financial support from Belspo for contract PRODEX GAIA-DPAC. I am indebted to Rosine Lallement for providing me with the FEROS spectrum of $\sigma$ Lupi. I also wish to thank Huib Henrichs and Gregg Wade for providing their results for $\sigma$ Lup prior to publication, as well as Swetlana Hubrig for valuable comments on the first draft of this paper. Yassine Damerdji and Andrea Miglio are thanked for their help with the preparation of the figure.

\bibliography{morel}

\end{document}